\documentclass[aps,prb,amsfonts,amssymb,twocolumn,amsmath,preprintnumbers,nofootinbib,floatfix,
showpacs]{revtex4-1}
\usepackage[dvips]{graphics}
\usepackage{graphicx}
\usepackage{bm}
\begin{document}

\title{Order parameter sign-reversal near $s_\pm$-superconductor surface}
\author{A. M. Bobkov}
\affiliation{Institute of Solid State Physics, Chernogolovka,
Moscow reg., 142432 Russia}
\author{I. V. Bobkova}
\affiliation{Institute of Solid State Physics, Chernogolovka,
Moscow reg., 142432 Russia}

\date{\today}

\begin{abstract}
The superconducting order parameter and LDOS spectra near an impenetrable surface are studied
on the basis of selfconsistent calculations for a two band superconductor with nodeless 
extended s±-wave order parameter symmetry, as possibly realized in Fe-based high-temperature superconductors. 
It is found that for a wide range of parameters the spatial behavior of the order parameter at a surface
is not reduced to a trivial suppression. If the interband scattering at a surface 
is of the order of the intraband one or dominates it, it can be energetically favorable to change the symmetry 
of the superconducting state near the surface from $s_{\pm}$ to conventional $s$-wave. 
The range of existing this surface conventional superconductivity
is very sensitive to the relative values of interband and intraband pairing potentials. It is shown 
that the LDOS spectra near the surface can qualitatively differ upon calculating with and without 
taking into account the selfconsistency of the order parameter.
\end{abstract}


\maketitle

The discovery of a new family of iron-based high-temperature superconductors with
distinct multi-orbital band structure \cite{bruning08,zhao08,pourovskii08} 
has renewed interest to the problem of multi-band superconductivity, firstly
discussed fifty years ago \cite{suhl59,moskalenko59}. It was proposed 
theoretically \cite{mazin08,kuroki08} that the Fe-based superconductors represent the first example of
multigap superconductivity with a phase difference between the superconducting
condensates belonging to different bands. This state was discussed previously 
\cite{golubov97,artenberg99}, but not yet observed in nature. In the most simple case there is 
the phase difference $\pi$ between the
superconducting condensates arising on the hole Fermi surfaces around 
$\Gamma$ point and the electron Fermi surfaces
around M point. This so-called $s_{\pm}$ (or extended $s$-wave) state
has been favored by a variety of models within random phase approximation
(RPA) \cite{kuroki08,cvetkovic09_1,graser09} and renormalization group techniques 
\cite{chubukov08,wang09,cvetkovic09_2}. Currently
the $s_{\pm}$-state is viewed to be the most plausible candidate for the role
of the superconducting order parameter in these compounds.

Surface and interface phenomena in $s_{\pm}$-superconductors have attracted
considerable recent attention. The formation
of bound states at a free surface of an $s_{\pm}$-superconductor \cite{ghaemi09,nagai09_1,onari09,nagai09_2},
at an S$_\pm$/N \cite{choi08,linder09,golubov09,sperstad09}, an N/S/S$_\pm$ junction \cite{feng09}
and at Josephson junctions including $s_{\pm}$-superconductors \cite{tsai09,sperstad09} was investigated theoretically.
In particular, the finite energy subgap bound states (depending on the interface parameters) were found 
and their influence on the conductance spectra and Josephson current was investigated.

However almost all these calculations (except for a few numerical results \cite{choi08}) assume non-selfconsistent
superconducting order parameter (OP). In the present paper we focus on the study of the OP at a surface of 
$s_{\pm}$-superconductor. We have found that for a wide range of parameters the spatial behavior of the OP 
at a surface can not be reduced to a trivial suppression. If the interband scattering at a surface $R_{12}$ 
is of the order of the intraband one $R_0$ or dominates it, it can be energetically favorable 
to change the symmetry of the superconducting state near the 
surface from $s_{\pm}$ to conventional $s$-wave. The range of existing this surface conventional superconductivity
is very sensitive to the relative values of interband and intraband pairing potentials. We demonstrate that
the selfconsistent OP behavior affects the surface LDOS profiles, and,
consequently, should be taking into account when interpreting experimental results. It is worth to note here
that, while there is a wide parameter range of
existing complex OP at the surface region \cite{bobkov}, in this paper we only
discuss the case when the surface OP is of conventional $s$-wave type. 

We consider an impenetrable surface of a clean two-band superconductor. The OP
is assumed to be of $s_\pm$-symmetry in the bulk of the superconductor, that is the phase difference
between the OP's in the two bands (called 1 and 2) is $\pi$. It is supposed that an incoming quasiparticle
from band 1,2 can be scattered by the surface as into the same band (intraband scattering), so as into
the other band (interband scattering). 

We make use of the quasiclassical theory of superconductivity, where all the relevant physical
information is contained in the quasiclassical Green function
$\hat g_i(\varepsilon, \bm p_f, x)$ for a given quasiparticle trajectory. 
Here $\varepsilon$ is the quasiparticle energy measured
from the chemical potential, $\bm p_f$ is the 
momentum on the Fermi surface (that can have several
branches), corresponding to the considered trajectory,
$x$ is the spatial coordinate along the normal to the surface and
$i=1,2$ is the band index. Quasiclassical Green function
is a $2 \times 2$ matrix in  particle-hole space, that is denoted by the symbol $\hat ~$. The equation
of motion for $\hat g_i(\varepsilon, \bm p_f, x)$ is the Eilenberger equation
subject to the normalization condition \cite{larkin69,eilenberger68}. For superconductivity of $s_\pm$-type,
when the pairing of electrons from different bands is absent, the Eilenberger equations
corresponding to the bands 1 and 2 are independent. The trajectories belonging to the different bands 
can only be entangled by the surface, which enters the quasiclassical theory in the form of effective
boundary conditions connecting the
incident and outgoing trajectories. 

However, owing to the normalization
condition for the quasiclassical propagator, the boundary
conditions for the quasiclassical Green functions are formulated 
as non-linear equations \cite{shelankov84,zaitsev84,millis88}.
Furthermore, they contain unphysical, spurious solutions, so their practical use is limited.
For this reason in the present work we make use of the quasiclassical formalism
in term of so-called Riccati amplitudes \cite{shelankov80,eschrig00}, that allows an explicit formulation
of boundary conditions \cite{eschrig00,shelankov00,fogelstrom00,zhao04,eschrig09}. The retarded Green function 
$\hat g_i(\varepsilon, \bm p_f, x)$, 
which is enough for a complete description of an equilibrium system, can be parametrized via two
Riccati amplitudes (coherence functions) $\gamma_i(\varepsilon, \bm p_f, x)$ and 
$\tilde \gamma_i(\varepsilon, \bm p_f, x)$ (in the present paper we follow 
the notations of Refs.~\cite{eschrig00,eschrig09}).  
The coherence functions obey the Riccati-type transport equations.
In the considered here case of two-band clean $s_\pm$-superconductor the equations for the two bands are independent
and read as follows
\begin{equation}
iv_{ix}\partial_x \gamma_i+2\varepsilon \gamma_i =-\Delta_i^* \gamma_i^2-\Delta_i
\label{gamma}
\enspace ,
\end{equation}
\begin{equation}
\tilde \gamma_i (\varepsilon, \bm p_f, x)=\gamma_i^*(-\varepsilon, -\bm p_f, x)
\label{tilde_gamma}
\enspace .
\end{equation}
Here $v_{ix}$ is the normal to the surface Fermi velocity component for the quasiparticle belonging to band $i$.
$\Delta_i$ stands for the OP in the $i$-th band, which should be found self-consistently.

Let us suppose that the surface is located at $x=0$ and the superconductor occupies the halfspace $x>0$. 
For the sake of simplicity we assume that the surface is atomically clean and, consequently,
conserves parallel momentum component. Then there are four quasiparticle trajectories, which are involved
in each surface scattering event. These are two incoming trajectories belonging to the bands 1,2 (with $v_{ix}<0$)
and two outgoing ones (with $v_{ix}>0$). It can be shown \cite{eschrig00,eschrig09} that the coherence function
$\gamma_i(\varepsilon, \bm p_f, x)$, corresponding to the incoming trajectory 
can be unambiguously calculated making use of Eq.~(\ref{gamma}) 
up to the surface starting from its asymptotic value in the bulk
\begin{equation}
\gamma_i^{b}=-\frac{\Delta_i^b{\rm sgn}\varepsilon}{|\varepsilon|+\sqrt{(\varepsilon+i\delta)^2-{\Delta_i^b}^2}}
\label{gamma_bulk}
\enspace ,
\end{equation}
where $\Delta_i^b$ is the bulk value of the OP in the appropriate band, $\delta>0$ is an infinitesimal. 
As for the coherence function
$\tilde \gamma_i(\varepsilon, \bm p_f, x)$, it is determined unambiguously by the asymptotic conditions for
the outgoing trajectories and can be obtained according to Eqs.~(\ref{gamma}),(\ref{tilde_gamma}).

Otherwise, the coherence functions $\gamma_i(\varepsilon, \bm p_f, x)$ for the outgoing trajectories
and, correspondingly, $\tilde \gamma_i(\varepsilon, \bm p_f, x)$ for the incoming ones should be calculated
from Eq.~(\ref{gamma}) supplemented by the boundary conditions at the surface and Eq.~(\ref{tilde_gamma}). 
The surface is described by
the normal state scattering matrix for particle-like excitations, denoted by $S$ and  
for hole-like excitations, denoted by $\widetilde S$, that connect outgoing with incoming quasiparticles.
The scattering matrix $S$ have elements $S_{\bm k_i\bm p_j}$, 
which connect outgoing quasiparticles from band $i$ with momentum $\bm k_i$ to the incoming ones 
belonging to band $j$ with momentum $\bm p_j$.
Here and below all the momenta corresponding to the incoming trajectories are denoted by letter $\bm p$
and all the momenta for the outgoing quasiparticles are denoted by $\bm k$. For the model we consider $S$
is a $2 \times 2$-matrix (for the particular value of the momentum parallel to the surface) in the trajectory space.
It obeys the unitary condition $SS^\dagger=1$ and without loss of generality 
can be parameterized by three quantities $R_{12}$, $\Theta$ and $\alpha$ as follows
\begin{equation}
\left(
\begin{array}{cc}
S_{\bm k_1\bm p_1} & S_{\bm k_1\bm p_2} \\
S_{\bm k_2\bm p_1} & S_{\bm k_2\bm p_2} 
\end{array}
\right)=
\left(
\begin{array}{cc}
\sqrt R_0 e^{i\Theta} & i \alpha \sqrt{R_{12}} \\
i \alpha \sqrt{R_{12}} & \sqrt R_0 e^{-i\Theta}
\end{array}
\right)
\label{S_electron}
\enspace ,
\end{equation}
where $R_0$ and $R_{12}$ are coefficients 
of intraband and interband reflection, respectively. They obey the constraint $R_0+R_{12}=1$. 
The phase factors $\alpha=\pm1$ and $\Theta$ appear to be unimportant for further consideration.
While in general the scattering matrix elements are functions of the momentum parallel to the surface 
$\bm p_{||}$, we disregard this dependence in order to simplify the analysis. 
The scattering matrix $\widetilde S$ for hole-like excitations are connected to $S$ by the relation
$\widetilde S(\bm p_{||})=S^{tr}(-\bm p_{||})$. In the absence of spin-orbit interaction the $S$-matrix
elements are only functions of $|\bm p_{||}|$, that is in the case we consider $\widetilde S=S$.

From the general boundary conditions \cite{eschrig09}, which are also valid
for a multiband system, one can obtain the explicit values of the coherence functions
$\gamma_i (\varepsilon, \bm k, x=0)$ and $\tilde \gamma_i(\varepsilon, \bm p, x=0)$
via the scattering matrix elements and the values of the coherence functions
$\gamma_i (\varepsilon, \bm p, x=0)$ and $\tilde \gamma_i(\varepsilon, \bm k, x=0)$
at the surface. They read as follows
\begin{equation}
\gamma_{1\bm k}=R_0 \gamma_{1\bm p}+R_{12} \gamma_{2\bm p}-\frac{R_0R_{12}\tilde \gamma_{2\bm k}
(\gamma_{1\bm p}-\gamma_{2\bm p})^2}
{1+\tilde \gamma_{2\bm k}\left( R_{12} \gamma_{1\bm p}+R_0 \gamma_{2\bm p} \right)}
\label{gamma_surface}
\enspace ,
\end{equation}
\begin{equation}
\tilde \gamma_{1\bm p}=R_0 \tilde \gamma_{1\bm k}+R_{12} \tilde \gamma_{2\bm k}-\frac{R_0R_{12}
\gamma_{2\bm p}(\tilde \gamma_{1\bm k}-\tilde \gamma_{2\bm k})^2}
{1+\gamma_{2\bm p}\left( R_{12} \tilde \gamma_{1\bm k}+R_0 \tilde \gamma_{2\bm k} \right)}
\label{tilde_gamma_surface}
\enspace .
\end{equation}
Here the arguments $(\varepsilon,x=0)$ of all the coherence functions are omitted for brevity, 
$\gamma_{i\bm p} \equiv \gamma_i(\bm p)$ and $\tilde \gamma_{i\bm p} \equiv \tilde \gamma_i(\bm p)$
and the analogous notations are used for $\gamma_i(\bm k)$ and $\tilde \gamma_i(\bm k)$.
Quantities $\gamma_{i\bm p}$ and $\tilde \gamma_{i\bm k}$, entering Eqs.~(\ref{gamma_surface}) 
and (\ref{tilde_gamma_surface}) are to be calculated from Eqs.~(\ref{gamma}) and (\ref{tilde_gamma}) 
supplemented by the appropriate asymptotic condition. 
The coherence functions $\gamma_{2\bm k}$ and $\tilde \gamma_{2\bm p}$ are obtained by the interchanging 
$1 \leftrightarrow 2$ in all the coherence function band indices at the right-hand side of 
Eqs.~(\ref{gamma_surface}) and (\ref{tilde_gamma_surface}), respectively.

Now, substituting the coherence functions into the self-consistency equation
\begin{equation}
\Delta_i(x)=-T\sum \limits_{\varepsilon_n,j}\lambda_{ij}\left \langle 
\frac{-2 i \pi \gamma_{j\bm p_f}}{1+\gamma_{j\bm p_f}\tilde \gamma_{j \bm p_f}} \right \rangle_{\bm p_f}
\label{self_consistency}
\enspace ,
\end{equation}
we iterate system (\ref{gamma})-(\ref{gamma_bulk}), (\ref{gamma_surface})-(\ref{self_consistency})
until it converges. In Eq.~(\ref{self_consistency}) $\lambda_{ii}<0$ is the dimensionless 
pairing potential for band $i$ and $\lambda_{12}=\lambda_{21}$ is the dimensionless interband
pair-scattering potential. We choose $\lambda_{12}>0$, which stabilizes $s_\pm$ OP in the bulk.
The Matsubara frequencies $\varepsilon_n$ enter the coherence functions via the substitution 
$\varepsilon+i\delta \to i \varepsilon_n$. $\langle ... \rangle_{\bm p_f}$ means the anomalous Green function
averaged over the entire Fermi surface, that is $\bm p_f$ incorporates as the incoming trajectories 
$\bm p$, so as the outgoing ones $\bm k$. For concreteness  we suppose the Fermi surface to
be cylindrical for the each band. However, our results do not qualitatively sensitive to this assumption.   

\begin{figure}[!tbh]
  \centerline{\includegraphics[clip=true,width=3.1in]{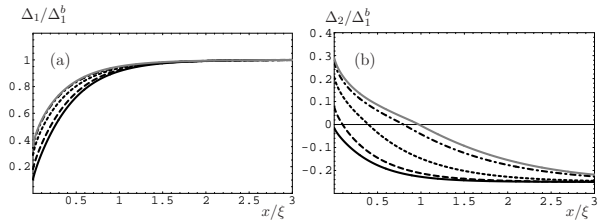}}
   \caption{Spatial profiles of the OP corresponding to band $1$ (panel(a))
and band $2$ (panel(b)) near the surface ($x=0$) for different values of interband pair scattering. $R_{12}=1$, 
$\Delta_2^b=-0.25 \Delta_1^b$, temperature $T=0.3\Delta_1^b$.
The values of intraband pairing potentials $\lambda_{ii}$ are adjusted
to keep $\Delta_1^b$ and $\Delta_2^b$ unchanged upon varying $\lambda_{12}$. The particular values
of coupling constants are the following: $(\lambda_{12},\lambda_{11},\lambda_{22})=(0.06,-0.2607,-0.0327)$
for black solid curve, $(0.03,-0.2695,-0.1353)$ for dashed curve, $(0.01,-0.2753,-0.2036)$
for dotted curve, $(0.004,-0.2771,-0.2241)$ for dashed-dotted curve and $(0.002,-0.2777,-0.2310)$
for gray solid curve. The superconducting coherence length 
$\xi=v_{1f}/\Delta_1^b$.}
\label{Delta_lambda_inter}
\end{figure} 

\begin{figure}[!tbh]
  \centerline{\includegraphics[clip=true,width=3.1in]{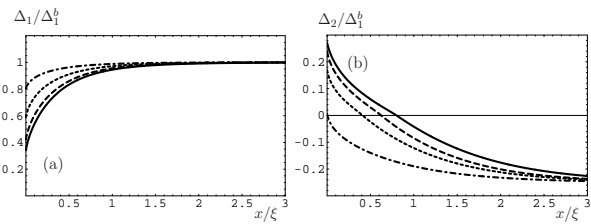}}
   \caption{Spatial profiles of the OP near the surface in band $1$ (panel (a))
and in band $2$ (panel (b)) for different values of interband surface scattering $R_{12}$. 
$\Delta_2^b=-0.25 \Delta_1^b$, $T=0.3\Delta_1^b$, 
$(\lambda_{12},\lambda_{11},\lambda_{22})=(0.004,-0.2771,-0.2241)$.
$R_{12}=1$ for black solid curve, $0.8$ for dashed curve, $0.5$ for dotted curve and $0.2$ for dashed-dotted curve.}
\label{Delta_intraband_scattering}
\end{figure}

\begin{figure}[!tbh]
  \centerline{\includegraphics[clip=true,width=3.1in]{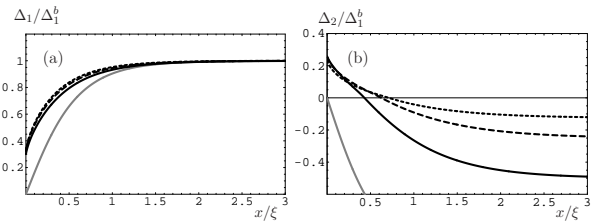}}
   \caption{Spatial profiles of the OP near the surface in band $1$ (panel (a))
and in band $2$ (panel (b)) for different values of the ratio $\Delta_2^b/\Delta_1^b$. 
$T=0.3\Delta_1^b$, $\lambda_{12}=0.005$, $R_{12}=1$. $\Delta_2^b/\Delta_1^b=-1$ ($\lambda_{11}=\lambda_{22}=0.2732$) 
for gray solid curve, $\Delta_2^b/\Delta_1^b=-0.5$ ($\lambda_{11}=0.2754,
\lambda_{22}=0.2393$) for black solid curve, $\Delta_2^b/\Delta_1^b=-0.25$ ($\lambda_{11}=0.2768,
\lambda_{22}=0.2207$) for dashed curve and $\Delta_2^b/\Delta_1^b=-0.125$ ($\lambda_{11}=0.2775,
\lambda_{22}=0.2011$) for dotted curve.}
\label{Delta_lambda_intra}
\end{figure}

The spatial profiles of the OP calculated according to the described above technique,
are represented in Figs.~\ref{Delta_lambda_inter}-\ref{Delta_lambda_intra}. We assume that
in the bulk $|\Delta_1^b|>|\Delta_2^b|$. Panels (a) of Figs.~\ref{Delta_lambda_inter}-\ref{Delta_lambda_intra}
demonstrate the spatial OP profiles for band $1$, while panels (b) correspond to band $2$. 

Fig.~{\ref{Delta_lambda_inter}} shows the dependence of the effect on the interband pair-scattering value 
$\lambda_{12}$. While the larger OP (in band $1$) is simply suppressed near the surface and the   
magnitude of the suppression is only slightly sensitive to $\lambda_{12}$ (at least, in the range we consider),
the smaller OP (in band $2$) reverses its sign near the surface. Thus, there is a surface region, which size
is comparable to the superconducting coherence length $\xi$, where $s_\pm$-superconductivity is superseded
by the conventional one. The reason for this OP sign reversal is the interband surface scattering $R_{12}$,
because it is energetically more favorable to minimize the OP gradient term along the quasiparticle trajectory.
It is worth to note here somewhat similar effect of the OP sign reversal for the smaller gap 
due to magnetic impurities \cite{golubov97}.

Fig.~{\ref{Delta_lambda_inter}} demonstrates that the weaker the interband pair scattering potential 
the wider the region, where the conventional $s$-wave superconductivy exists. It can be qualitatively understood
on the basis of the self-consistency equation (\ref{self_consistency}): the phase of the smaller OP
is determined to a great extent by the phase of the dominant OP via the term $\lambda_{12}$. 
The weaker this connection the less energy cost to reverse the phase. As it is seen from the caption to
Fig.~{\ref{Delta_lambda_inter}}, in the particular calculations we take $|\lambda_{12}|<<|\lambda_{11}|$.
Such a choice of parameters is consistent with the experimental estimates of the coupling constants
for FeSe \cite{khasanov09}.

The curves represented in Fig.~\ref{Delta_lambda_inter} are calculated under the assumption of
purely interband surface scattering $R_{12}=1$, when the OP sign reversal is strongest.
In order to investigate the effect in the more realistic situation one needs to take into account
intraband scattering $R_0$. The corresponding results are demonstrated in Fig.~\ref{Delta_intraband_scattering}.
It is seen that the region of reversed OP existence shrinks with increasing of $R_0$. However the effect
remains to be pronounced even if the portion of intraband scattering exceeds $50\%$. Thus, we believe that
the self-consistent OP treatment is essential for polycrystalline samples, for example, upon analysing
spectroscopic data. 

The considerable region of conventional $s$-wave surface superconductivity can only occur
if the bulk OP's in the two bands essentially differ in magnitude. In case if $|\Delta_2^b|$
approaches to $|\Delta_1^b|$, the OP phase reversal region shrinks and, finally, the surface OP
behavior reduces to the trivial suppression for the two bands if $|\Delta_1^b|=|\Delta_2^b|$.
Otherwise, upon decreasing the ratio $|\Delta_2^b|/|\Delta_1^b|$ the surface region gets wider
until its width saturates at some value of the ratio ($\sim 1/4$ for the considered case). This
is illustrated in Fig.~\ref{Delta_lambda_intra}.

\begin{figure}[!tbh]
  \centerline{\includegraphics[clip=true,width=3.3in]{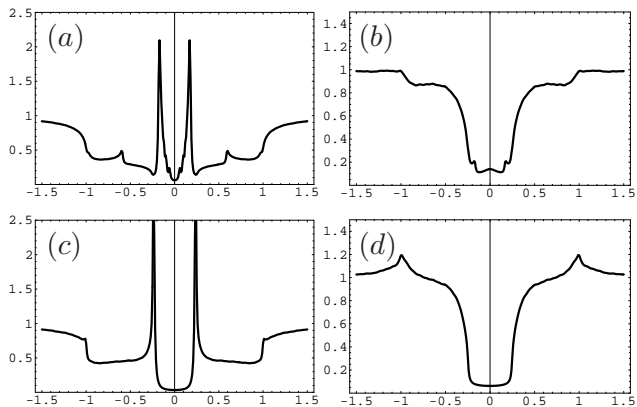}}
   \caption{LDOS as a function of quasiparticle energy calculated at $x=0$. The energy
is measured in units of $\Delta_1^b$. Left column
represents LDOS for band $1$, while right column corresponds to band $2$. Upper row 
demonstrates the results of selfconsistent calculations and lower one shows the LDOS
assuming non-selfconsistent OP. $(\lambda_{12},\lambda_{11},\lambda_{22})=(0.004,-0.2771,-0.2241)$,
$\Delta_2^b=-0.25 \Delta_1^b$, $T=0.3\Delta_1^b$, $R_{12}=0.5$.}
\label{LDOS_R_05}
\end{figure}

\begin{figure}[!tbh]
  \centerline{\includegraphics[clip=true,width=3.3in]{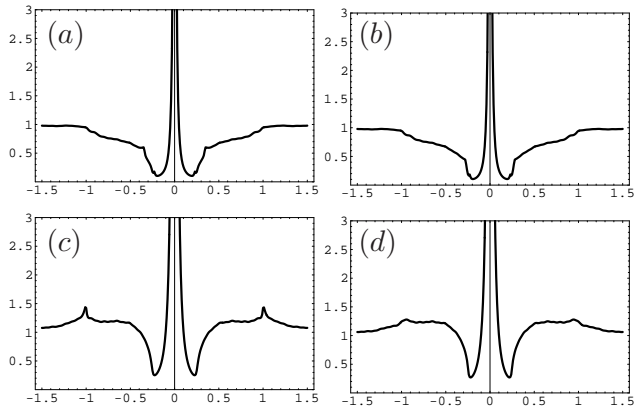}}
   \caption{The same as in Fig.~\ref{LDOS_R_05}, but corresponding to $R_{12}=1$.}
\label{LDOS_R_1}
\end{figure}

Now we discuss how the described above OP spatial behavior affects the local density of states (LDOS)
near the surface. LDOS $\rho_i$ corresponding to band $i$ is calculated via the coherence functions as follows
\begin{equation}
\rho_i(x,\varepsilon)={\rm Re} \left \langle 
\frac{1-\gamma_{i\bm p_f}\tilde \gamma_{i \bm p_f}}
{1+\gamma_{i\bm p_f}\tilde \gamma_{i \bm p_f}} \right \rangle_{\bm p_f}
\label{LDOS}
\enspace .
\end{equation}

Figs.~\ref{LDOS_R_05} and \ref{LDOS_R_1} represent the LDOS at the surface ($x=0$)
as a function of quasiparticle energy. Left and right columns of the Figures demonstrate the LDOS 
for bands $1$ and $2$ separately. Upper row of each Figure shows LDOS plots, calculated taking into
account the selfconsistent OP behavior. It should be compared to the lower row, where LDOS is plotted 
for the same parameters, but for the non-selfconsistent OP equal to its bulk value. The results represented
in Fig.~\ref{LDOS_R_05} correspond to $R_{12}=0.5$, while Fig.~\ref{LDOS_R_1} illustrates the case
of purely interband scattering $R_{12}=1$. 

As it was already discussed in the literature, if $R_{12} \neq 0$ there are surface 
bound states in the system, which manifest themselves as well-pronounced peaks in the LDOS.
At $R_{12} \to 1$ the bound state energies tend to zero, what can be clearly seen from the corresponding
LDOS plots. For this case the LDOS is dominated by very strong zero-energy peak and the differences
between selfconsistent and non-selfconsistent plots are not qualitative. It can be only noted 
that the features corresponding to gap edges (clearly seen at least in panel (c) of Fig.~\ref{LDOS_R_1})
are washed out under selfconsistent calculation. At the same time for the intermediate value of interband
scattering there are qualitative differencies between selfconsistent and non-selfconsistent results 
(see Fig.~\ref{LDOS_R_05}).
They can be summarize as follows: (i) while in the non-selfconsistent picture the bound state peaks are divided
by the clearly defined gap, selfconsistency results in transforming this inner gap into "V"-shaped behavior,
which is known to be more typical for the superconductors with OP nodes at the Fermi surface. We believe
that this observation can be essential for interpreting experimental data. (ii) additional
features (small peaks) appear in the subgap region upon taking into account selfconsistency.
It is worth to note here that, in contrast to the interface OP behavior, the shape of LDOS profiles 
can be quite sensitive to the details of the microscopic model describing the interface,
in particular to the concrete dependence of the scattering matrix elements on $\bm p_{||}$. 
However, if the particular microscopic scattering matrix model leads to the existence of an inner gap
in the LDOS, it is inevitably transformed into "V"-shaped behavior under selfconsistent calculation, 
as it is demonstrated above. This fact is a consequence of 
spatial line of OP nodes appearing if the OP sign reversal takes place at the surface.

In summary, for a two band $s_\pm$-superconductor we have theoretically investigated 
the behavior of the OP at a specular reflecting surface. It is found that if the interband 
surface scattering is of the order of the intraband one or dominates it, the OP belonging to the band 
with smaller OP absolute value in the bulk reverses its sign at a surface region, thus giving rise to
conventional $s$-wave surface superconductivity. This region of reversal sign OP is maximal for purely
interband surface scattering and shrinks upon its diminishing. The effect is quite sensitive to the ratio
between intraband and interband superconducting coupling constants and is more pronounced for smaller
interband coupling constant. It is also shown that if the OP sign reversal
takes place, it results in qualitative changes in the LDOS spectra near the surface.

The support by RFBR Grant 09-02-00779 and the programs of Physical Science Division
of RAS is acknowledged. A.M.B. was also supported by the Russian
Science Support Foundation.


\end{document}